\documentclass[11pt,a4paper]{article}
\ifdefined\directlua\else
  \pdfoutput=1
  \usepackage[T1]{fontenc}
  \usepackage[utf8]{inputenc}
\fi
\usepackage{lmodern}
\usepackage[english]{babel}
\usepackage{amsmath,amssymb,amsthm,mathtools}
\usepackage{booktabs,array,tabularx}
\usepackage{microtype}
\usepackage[a4paper,margin=25mm]{geometry}
\usepackage[hidelinks]{hyperref}
\usepackage{enumerate}
\usepackage{xcolor}
\usepackage{fancyhdr}
\usepackage{flafter}
\usepackage{tikz}
\usetikzlibrary{positioning,arrows.meta}

\makeatletter
\g@addto@macro\@listi{\setlength{\itemsep}{1pt}\setlength{\parsep}{0pt}\setlength{\topsep}{3pt}}
\makeatother

\newtheorem{theorem}{Theorem}[section]
\newtheorem{lemma}[theorem]{Lemma}
\newtheorem{proposition}[theorem]{Proposition}
\newtheorem{corollary}[theorem]{Corollary}
\theoremstyle{definition}
\newtheorem{definition}[theorem]{Definition}
\theoremstyle{remark}
\newtheorem{remark}[theorem]{Remark}

\newcommand{\M}{\mathsf M}
\newcommand{\R}{\mathsf R}
\newcommand{\UU}{\operatorname{UNO}}
\newcommand{\rd}{\operatorname{rd}}
\newcommand{\Off}{\operatorname{off}}
\newcommand{\PRE}{\mathsf{PRE}}
\newcommand{\TEST}{\mathsf{TEST}}
\newcommand{\GIVE}{\mathsf{GIVE}}
\newcommand{\FLAG}{\mathsf{FLAG}}
\newcommand{\CORE}{\mathsf{CORE}}
\newcommand{\HALT}{\mathsf{HALT}}
\newcommand{\INC}{\mathsf{INC}}
\newcommand{\DEC}{\mathsf{DEC}}
\newcommand{\bb}[1]{\texttt{#1}}

\title{Turing Completeness of Greedy UNO\\[-1pt]
\large with an Ultimately Periodic Infinite Deck}
\author{T. V. Kuptsov\thanks{Faculty of Mechanics and Mathematics, Lomonosov Moscow State University, Moscow, Russia.}
\and G. A. Khaziev\thanks{Institute for Information Transmission Problems, Russian Academy of Sciences (Kharkevich Institute), Moscow, Russia; Central University, Moscow, Russia}}
\date{16 November 2026}

\begin{document}
\maketitle

\begin{abstract}
We encode an arbitrary two-counter Minsky machine as a two-player UNO
round with an ultimately periodic infinite draw pile $uw^\omega$.
The construction uses a fixed $17$-card prefix and a period of length
quadratic in the number of instructions. When both players follow the
same fixed greedy policy, the round terminates exactly when the machine
halts, and every available hand play has a unique card type. Consequently,
greedy termination and victory by a designated player---as well as
winning-strategy existence under forced-play rules without tie-breaking---are
$\Sigma^0_1$-complete. Under official voluntary-draw rules with unrestricted
legal choices, however, the existence of a finite play is decidable (and
primitive recursive) via reachability in fixed-dimension vector addition
systems with states (VASS). Finally, the maximal duration of a finite greedy
play has no total computable upper bound in the input length.
\end{abstract}



\section{Introduction}\label{sec:introduction}

Finite models of UNO have been studied in algorithmic game theory. Demaine et al.\ proved NP-completeness for a one-player model and polynomial-time solvability for a particular uncooperative two-player model~\cite{demaine2014}. Other card games can simulate arbitrary computation. For \emph{Magic: The Gathering}, undecidability holds even for positions without further meaningful move choices~\cite{churchill2020}, and stronger arithmetic complexity is established in~\cite{biderman2020}. A result for \emph{Yu-Gi-Oh!} concerns winning strategies~\cite{yugioh2026}. These games use richer card-specific effects than the color-or-symbol matching, Skip, and Draw Two operations considered here. Our infinite-deck model differs from the finite models of~\cite{demaine2014}; the complexity results concern different decision problems.

The draw pile in our model is infinite but has a finite description: a finite prefix followed by repetitions of a nonempty finite period. Each hand is finite at every finite time, although hand sizes have no a priori bound. Multiplicities of a few card types serve as counters, while the repeating draw pile acts as a read-only program controlling their updates.

The main theorem concerns the \emph{unique play induced by the same fixed computable policy for both players}. The rules permit a player to draw instead of playing a compatible card, as in the published Mattel rules~\cite{mattelrules}. The greedy policy always plays an available compatible card, and in the constructed instances every available hand play has a unique card type, so the result extends directly to two-player forced-play rules without a tie-breaking order (Corollary~\ref{cor:forced}). Thus \emph{Turing completeness} refers to these induced dynamics; the theorem makes no claim about optimal play in ordinary finite-deck UNO.

This distinction between mandatory play and voluntary drawing changes decidability. Mandatory play makes the absence of a card type observable through a forced draw. Under unrestricted legal choices, a draw does not certify absence, and existential termination can instead be expressed as VASS reachability in fixed dimension. Theorem~\ref{thm:vass} establishes the resulting decidability and primitive-recursive upper bound.

Section~\ref{sec:model} defines the game and the decision problems, and Section~\ref{sec:main} states the main theorem and its complexity consequences. Sections~\ref{sec:encoding}--\ref{sec:simulation} present and verify the simulation. Section~\ref{sec:duration} studies maximal finite duration, and Section~\ref{sec:unrestricted} treats unrestricted legal play. The appendices collect operational details, transition certificates, additional traces, size bounds, and computational checks.

\section{Game model and decision problems}\label{sec:model}

\subsection{Cards and turn rules}

Let
\[
\Gamma=\{r,y,g,b\}\times\bigl(\{0,1,\ldots,9\}\cup\{S,D\}\bigr).
\]
The first coordinate is a color and the second is a numeral, $S$ (Skip), or $D$ (Draw Two). We denote $(g,0)$ by $\bb{g0}$, and so on. Two cards are \emph{compatible} if their colors or symbols coincide. A hand is a finite multiset over $\Gamma$; a card \emph{type} means one element of $\Gamma$, not an individual physical copy.

There are two players, $\M$ and $\R$. The draw pile is an infinite word $d_0d_1\ldots\in\Gamma^\omega$. Initially, $\R$ deals seven cards to each player in alternation, beginning with $\M$. The fifteenth card $d_{14}$ becomes the initial top discard (without triggering any Skip or Draw Two effect on the opening turn), and $\M$ takes the first turn.

On a turn, a player may either play a compatible card from their hand or draw one card. Drawing is allowed even when the hand contains a compatible card, and is mandatory when it does not. A compatible drawn card may be played immediately or retained; an incompatible drawn card must be retained. Retaining the card ends the turn. After drawing, the player may not play a different card from the hand on that turn.

Playing a numeral passes the turn to the opponent. A Skip card $S$ skips the opponent, so the same player takes the next turn. A Draw Two card $D$ forces the opponent to take the next two draw-pile cards \emph{without playing either}, skips that opponent's turn, and lets the player of $D$ move again. The round ends as soon as a player empties their hand, before any further card effects or draws.

We consider one round, without Wild or Reverse cards. UNO declarations are assumed to be timely, so declaration penalties do not occur. Scoring, subsequent rounds, recycling of discards, and shuffling are omitted. Drawing one card and the option to decline a playable card follow the manufacturer's rules~\cite{mattelrules}; the infinite draw pile and the stated omissions define the mathematical model studied here.

\subsection{Ultimately periodic draw piles}

The input is a pair of finite words $(u,w)$ over $\Gamma$, with $w\ne\varepsilon$, specifying the infinite draw pile $uw^\omega$. Its order is fixed and publicly known, and each type may occur arbitrarily many times. The model therefore describes a finite input with an unbounded supply of cards, rather than a physical UNO deck. The first fifteen cards determine the initial deal and discard even when $|u|<15$. The $k$-th card can be computed effectively from $(u,w,k)$.

\subsection{The greedy policy}

\begin{definition}[Fixed greedy policy]
Fix once and for all a linear order $\prec$ on $\Gamma$. The policy $\sigma$ plays the $\prec$-least compatible card type in the hand whenever such a type exists. Otherwise it draws one card and plays it immediately if compatible, retaining it otherwise. Both players follow $\sigma$ unless unrestricted or forced-play choices are explicitly specified.
\end{definition}

This policy is legal under the rules above. The tie-breaking order makes the play uniquely determined for every input. In the instances produced by the reduction, every available hand play has a unique type, so the order is never needed.

\subsection{Decision problems}

\begin{definition}
\textsc{Greedy-Uno-Termination} is the set of pairs $(u,w)$ of finite words over $\Gamma$, with $w\ne\varepsilon$, such that the game with draw pile $uw^\omega$, the standard seven-card initial deal, and both players following $\sigma$ terminates. \textsc{Greedy-Uno-R-Win} is the corresponding set of pairs for which $\R$ wins.
\end{definition}

Because the next draw-pile card is computable and both hands remain finite after any finite number of turns, the greedy play can be simulated for any prescribed finite number of turns.

\section{Two-counter machines and the main theorem}\label{sec:main}

A two-counter machine has nonnegative integer registers $A,B$ and a finite program $I_0,\ldots,I_{N-1}$ with labels in $\{0,\ldots,N-1\}$; the instruction set is
\[
\INC(X,t),\qquad \DEC(X,t_1,t_0),\qquad \HALT,
\qquad X\in\{A,B\}.
\]
An $\INC$ instruction increments $X$ and jumps to $t$. A $\DEC$ instruction decrements $X$ and jumps to $t_1$ when $X>0$, and otherwise jumps to $t_0$ without modifying $X$. A $\HALT$ instruction stops the machine. The initial state has instruction label $0$ and both counters equal to $0$. The halting problem for these machines is $\Sigma^0_1$-complete~\cite{minsky1967}.

\begin{theorem}[Main theorem]\label{thm:main}
There exist a fixed $u\in\Gamma^{17}$ and a computable mapping $\mathcal M\mapsto w_{\mathcal M}\in\Gamma^+$ such that, for every two-counter machine $\mathcal M$ with $N\ge1$ instructions, the unique greedy play on $u w_{\mathcal M}^\omega$ satisfies:
\begin{enumerate}[\upshape(i)]
\item At every reachable position the active player's compatible hand cards, if any, all have the same type.
\item The play terminates if and only if $\mathcal M$ halts; in that case $\R$ wins.
\item If $\mathcal M$ halts with registers $(a,b)$, then at the end $\M$ has exactly $a$ copies of $\bb{y2}$ and $b$ copies of $\bb{y3}$.
\item The period has length $|w_{\mathcal M}|=O(N^2)$, with an absolute implied constant.
\end{enumerate}
\end{theorem}

\begin{corollary}\label{cor:complete}
\textnormal{\textsc{Greedy-Uno-Termination}} and \textnormal{\textsc{Greedy-Uno-R-Win}} are $\Sigma^0_1$-complete. Hardness persists for a fixed $17$-card prefix and for instances whose greedy plays never offer more than one compatible hand-card type.
\end{corollary}
\begin{proof}
Both properties are recursively enumerable: simulate the unique greedy play and accept upon termination, or upon victory by $\R$, respectively. The many-one reduction from two-counter-machine halting furnished by Theorem~\ref{thm:main} establishes hardness. The uniqueness-of-type property is established for all constructed inputs; it need not itself be decidable.
\end{proof}

Because tie-breaking is never invoked on $u w_{\mathcal M}^\omega$, the reduction also settles strategic decision problems under the common \emph{forced-play} variant of UNO, in which drawing is permitted only when the hand contains no compatible card and a compatible drawn card must be played immediately, while a player holding multiple compatible card types may freely choose which compatible type to play.

\begin{corollary}[Forced-play game complexity]\label{cor:forced}
Under forced-play UNO on an ultimately periodic draw pile $uw^\omega$, existential termination, universal termination, and the existence of a strategy for $\R$ that guarantees victory in finitely many turns against all legal forced-play choices of $\M$ are all $\Sigma^0_1$-complete.
\end{corollary}
\begin{proof}
At every nonterminal forced-play state, the active player either has a unique draw action or chooses among at most $|\Gamma|=48$ compatible card types in the hand. Hence the forced-play game tree has finite out-degree and a computable transition relation. Existential termination is visibly recursively enumerable. By K\H{o}nig's lemma, every forced play terminates if and only if the game tree has a finite uniform depth bound, and $\R$ has a strategy guaranteeing finite-time victory if and only if $\R$ has a finite-depth winning strategy tree; both conditions can be verified by bounded search and are therefore in $\Sigma^0_1$. On $u w_{\mathcal M}^\omega$, Theorem~\ref{thm:main}(i) ensures that at most one legal forced-play action exists at every reachable state, so the entire forced-play game tree collapses to the single greedy play, yielding $\Sigma^0_1$-hardness for all three problems.
\end{proof}

\section{Encoding machine configurations}\label{sec:encoding}

\subsection{Card multiplicities and read boundaries}

Write
\[
J=\bb{r1},\qquad U_A=\bb{y2},\qquad U_B=\bb{y3},\qquad
E=\bb{y8},\qquad F=\bb{b9},\qquad s_A=2,\ s_B=3.
\]
An exponent $x^k$ on a card type denotes a multiset of $k$ copies; $E^+$ means an unspecified positive number of copies.

A \emph{read boundary} is a state immediately before the next draw-pile card is consumed, after all intervening hand plays prescribed by $\sigma$ have been completed. Write $\rd$ for the player who will consume that card. Such a boundary is reached after finitely many hand plays, since each reduces the total number of cards held. Appendix~\ref{app:semantics} gives the full state and microstep conventions.

\begin{definition}[Boundary configuration]\label{def:C}
For $j,a,b\ge0$, a configuration $\mathcal C(j,a,b)$ is a state with top discard $\bb{gS}$, player $\R$ about to consume the next pile card, and hands
\[
H_\R=\{F\},\qquad H_\M=J^{2j}U_A^aU_B^b E^+.
\]
The nonempty collection of reserve cards $E$ prevents $\M$ from winning accidentally.
\end{definition}

\subsection{Instruction addresses}

The period is partitioned into instruction \emph{segments}, $w=S_0S_1\cdots S_{N-1}$. An occurrence of a segment is \emph{active} if its input state has $j=0$ and \emph{passive} otherwise. The value $j$ counts the segments to be skipped before the next instruction is executed. To jump from instruction $i$ to instruction $t$, set
\[
\Off_i(t)=(t-i-1)\bmod N\in\{0,\ldots,N-1\}.
\]
For a $\DEC(X,t_1,t_0)$ instruction at label $i$, define
\begin{equation}\label{eq:oz}
o=\Off_i(t_1),\qquad
z=\Off_i(t_0)+N\,\mathbf1_{\{\Off_i(t_0)<o\}}.
\end{equation}
Thus $0\le o\le N-1$, $o\le z\le 2N-1$, and $z\equiv\Off_i(t_0)\pmod N$. The extra full period in the zero branch is used only to ensure $z-o\ge0$.

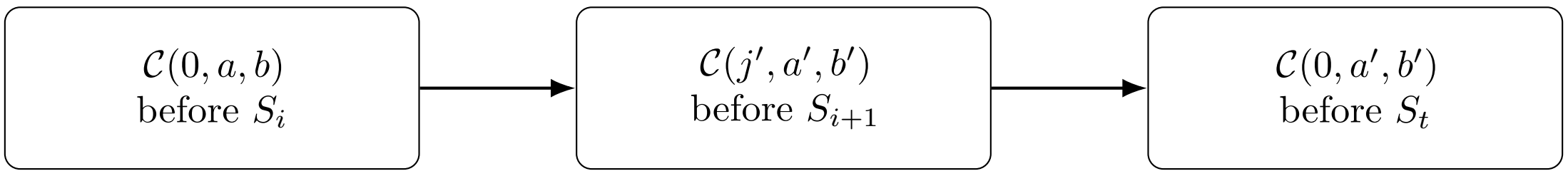
\begin{figure}[!htbp]
\centering
\begin{tikzpicture}[>=Latex,
state/.style={draw,rounded corners,align=center,font=\small,text width=37mm,
minimum height=15mm,inner sep=2pt}]
\node[state] (a) at (0,0) {$\mathcal C(0,a,b)$\\before $S_i$};
\node[state] (b) at (5.3,0) {$\mathcal C(j',a',b')$\\before $S_{i+1}$};
\node[state] (c) at (10.6,0) {$\mathcal C(0,a',b')$\\before $S_t$};
\draw[->,thick] (a.east) -- (b.west);
\draw[->,thick] (b.east) -- (c.west);
\node[below=8mm of b,align=center,font=\small]
{$j'\equiv t-i-1\pmod N,\qquad (i+1)+j'\equiv t\pmod N$};
\end{tikzpicture}
\caption{Instruction pointer: the first arrow is the active segment,
the second represents $j'$ passive segments (possibly zero).
Indices are understood modulo $N$; adding $N$ to $j'$ traverses one full period.}
\label{fig:jump}
\end{figure}

\section{Compiling instructions into deck segments}\label{sec:segments}

\subsection{Elementary blocks}

A juxtaposition of card symbols denotes concatenation; $\Pi^k$ is $k$ copies of the word $\Pi$. The constant blocks are
\begin{align*}
\TEST&=\bb{rS\ g0\ r0\ rS\ rS\ gS\ r4\ gS},\\
\GIVE&=\bb{g0\ g0\ yS\ g0},\\
\FLAG&=\bb{g4\ g4\ g0},\\
\Pi&=\bb{r1\ g0}.
\end{align*}
For a list $x_1,\ldots,x_m$ of card types, append one $E=\bb{y8}$ if $m$ is odd; the resulting even-length list has length $2h$. Define
\begin{equation}\label{eq:pre}
\PRE(x_1\cdots x_m)=\left(\prod_{k=1}^{h}\bb{gD}\,x_{2k-1}x_{2k}\right)\bb{gS}.
\end{equation}
The case $m=0$ gives $\PRE(\varepsilon)=\bb{gS}$. For $X\in\{A,B\}$, set
\[
c_X=\bb{g}s_X\in\{\bb{g2},\bb{g3}\},\qquad
\CORE_X=c_X\,\bb{gS\ gS\ g0\ gS\ gS}.
\]
Here $c_X$ is a \emph{single} card, not a concatenation of two symbols.

\subsection{Instruction segments and their specification}

For an instruction $I_i$, define its segment $S_i$ by one of the following cases:
\begin{align}
\INC(X,t):\quad S_i&=\TEST\,\PRE(J^oU_X)\,\GIVE\,\FLAG\,\Pi^o
\,c_X\,\bb{gS\ gS\ rS\ rS\ gS},\label{eq:inc}\\
\DEC(X,t_1,t_0):\quad S_i&=\TEST\,\PRE(J^{z+1})\,\GIVE\,\FLAG\,\Pi^o
\,\bb{g0}\,\CORE_X\,\bb{g0}\,\Pi^{z-o}
\,\bb{g1\ rS\ gS},\label{eq:dec}\\
\HALT:\quad S_i&=\TEST\,\FLAG\,\bb{g0\ gS\ bS\ gS\ g0\ gS}.
\label{eq:halt}
\end{align}
For the increment case $o=\Off_i(t)$; for the decrement case $o,z$ are as in~\eqref{eq:oz}. No exponent is negative. At most $16$ card types appear in the period, and at most $18$ appear in the entire construction including its initial prefix.

\begin{lemma}[Segment specification]\label{lem:spec}
Suppose $S_i$ is entered in $\mathcal C(j,a,b)$. If $j\ge1$, a segment of \emph{any} instruction type ends in $\mathcal C(j-1,a,b)$. If $j=0$, then:
\begin{enumerate}[\upshape(a)]
\item an $\INC(X,t)$ segment increments the encoded register $X$ and ends with $j'=\Off_i(t)$;
\item a $\DEC(X,t_1,t_0)$ segment, if $X>0$, decrements it and ends with $j'=o$, or, if $X=0$, leaves it unchanged and ends with $j'=z$;
\item an active $\HALT$ segment ends the game with $\R$ winning and leaves $U_A,U_B$ untouched.
\end{enumerate}
Every card of every completed nonterminal segment is consumed exactly once, and each card played from a hand has a uniquely determined type.
\end{lemma}

\section{Local gadget verification}\label{sec:gadgets}

\subsection{Boundary hand types and trace notation}

The gadgets are checked by complete action traces at read boundaries. In the compiled execution, the boundary hands consist of the following types:
\[
H_\M=J^m U_A^a U_B^b E^e(\bb{r4})^f,
\qquad H_\R=FJ^q(\bb{yS})^\eta,
\]
where $m,a,b,q\ge0$, $e\ge1$, and $f,\eta\in\{0,1\}$. Each lemma specifies which temporary cards may be present. In particular, the hands contain no green or symbol-$0$ cards at these boundaries. This restriction does not apply inside a gadget: for example, $\TEST$ temporarily leaves $\bb{g0}$ in $\R$'s hand.

For a player $P$, the notation $P:+x$ means ``draw $x$ and play it immediately'', $P:\langle x\rangle$ means ``draw and retain $x$'', and $P:-x$ means ``play $x$ from the hand without a draw''. Effects of $D$ are described separately. In the verification below, $F$ and at least one $E$ persist, so no terminal state interrupts the traces.

\subsection{Multiplicity invariance}

\begin{lemma}[Multiplicity invariance]\label{lem:copies}
Suppose two instances differ only in that one player has one extra copy of a type $x$. If the smaller-multiplicity instance retains at least one $x$ throughout a specified finite interval, and if the types of all playable cards from hands are unique, then both executions take the same sequence of actions and consume the same pile symbols during that interval; their multiplicities of $x$ continue to differ by one.
\end{lemma}
\begin{proof}
Induct on elementary actions. Whether a card type is available depends only on whether its multiplicity is zero or positive; by hypothesis these tests agree in the two executions. Therefore the chosen type, the compatibility of the newly drawn symbol, the effects of $S$ and $D$, and the subsequent active player agree. The difference in multiplicities is preserved. An early victory is excluded by the persistent reserve cards in the applications.
\end{proof}

\subsection{Test, loading, flag, and window}

\begin{lemma}[Test, loading, flag, and window]\label{lem:elementary}
In the states used by the compilation, the following hold without a choice of playable type:
\begin{enumerate}[\upshape(1)]
\item From $\mathcal C(j,a,b)$ the block $\TEST$ leaves the top $\bb{gS}$, $\rd=\R$, $H_\R=\{F\}$. If $j=0$, nothing changes in the hands; if $j\ge1$, two copies of $J$ are removed from $\M$ and one $\bb{r4}$ is added to $\M$.
\item When the top is $\bb{gS}$, $\rd=\R$, and $H_\R=\{F\}$ (no other cards in $\R$'s hand), the block $\PRE(x_1\cdots x_m)$ deposits all the listed $x_i$ and (if parity requires it) one $E$ in $\M$'s hand, returning to top $\bb{gS}$ and $\rd=\R$.
\item When the top is $\bb{gS}$, $\rd=\R$, and $\R$ has no green or zero-symbol card, $\GIVE$ stores exactly one $\bb{yS}$ in $\R$'s hand and ends at top $\bb{g0}$, $\rd=\R$.
\item Suppose the top is $\bb{g0}$ (or $\bb{gS}$ when $\eta=0$), $\rd=\R$, and the hands are $H_\M=J^m U_A^a U_B^b E^e(\bb{r4})^f$ and $H_\R=F(\bb{yS})^\eta$, for integers $m,a,b\ge0$, $e\ge1$, and $f,\eta\in\{0,1\}$. Then $\FLAG$ consumes the single $\bb{r4}$ from $\M$ if present. Afterwards the top is $\bb{g0}$ and $\rd=\R$ if the flag was present, $\rd=\M$ otherwise.
\item If the top is $\bb{g0}$, $\rd=P$, and both hands have no green or zero-symbol card, each $\Pi=\bb{r1\ g0}$ adds one $J$ to $P$'s hand without changing $\rd$ or the top. A single additional $\bb{g0}$ instead swaps $\rd$.
\end{enumerate}
\end{lemma}
\begin{proof}
For (1), the complete action traces, omitting only unchanged types $U_A,U_B,E,F$, are
\begin{center}\small
\begin{tabularx}{\linewidth}{@{}p{14mm}X@{}}
\toprule
Input & Actions in their exact order \\
\midrule
$j=0$ &
$\R:+\bb{rS}$, $\R:\langle\bb{g0}\rangle$,
$\M:+\bb{r0}$, $\R:-\bb{g0}$,
$\M:\langle\bb{rS}\rangle$, $\R:\langle\bb{rS}\rangle$,
$\M:+\bb{gS}$, $\M:-\bb{rS}$,
$\M:+\bb{r4}$, $\R:-\bb{rS}$, $\R:+\bb{gS}$. \\
$j\ge1$ &
$\R:+\bb{rS}$, $\R:\langle\bb{g0}\rangle$,
$\M:-J$, $\R:+\bb{r0}$, $\M:-J$,
$\R:+\bb{rS}$, $\R:+\bb{rS}$, $\R:+\bb{gS}$,
$\R:-\bb{g0}$, $\M:\langle\bb{r4}\rangle$,
$\R:+\bb{gS}$. \\
\bottomrule
\end{tabularx}
\end{center}
In both cases exactly the eight pile symbols of $\TEST$ are consumed. When $j=1$, the two original copies of $J$ disappear, while larger multiplicities leave a positive remainder. At the end, $\R$ holds only $F$.

For (2), $\R$ draws and plays each $\bb{gD}$ onto a green top. Each such play causes $\M$ to receive the next two symbols with no turn to play them. The closing $\bb{gS}$ is played by $\R$, who retains the turn. The case of an empty list also consists of a single green Skip.

For (3), the four pile symbols induce the sequence
$\R:+\bb{g0}$, $\M:+\bb{g0}$,
$\R:\langle\bb{yS}\rangle$, $\M:+\bb{g0}$.

For (4), when $\eta=0$ the hand $H_\R=\{F\}$ contains no Skip card, so top $\bb{gS}$ is a valid read boundary alongside top $\bb{g0}$. Player $\R$ first draws and plays $\bb{g4}$, which matches either $\bb{g0}$ or $\bb{gS}$ by color. When the flag exists, $\M:-\bb{r4}$, $\R:+\bb{g4}$, and $\M:+\bb{g0}$ follow. With no flag, $\M:+\bb{g4}$ and $\R:+\bb{g0}$ follow instead. These sequences establish the required reader changes and consume all three symbols.

For (5), $P$ first draws and keeps the incompatible $\bb{r1}$; the opponent then draws and plays $\bb{g0}$, so the reader returns to $P$. With a single $\bb{g0}$ and reader $P$, the precise trace is $P:+\bb{g0}$ followed by the other player taking the next draw. The boundary hand restrictions prevent an intervening play of another card type; see Appendix~\ref{app:traces} for all reader cases.
\end{proof}

\subsection{The decrement core}

\begin{lemma}[Decrement core]\label{lem:core}
Suppose the top is $\bb{g0}$ and the hands are $H_\M=J^m U_A^a U_B^b E^e$ and $H_\R=FJ^q\bb{yS}$, where $m,a,b,q\ge0$ and $e\ge1$. The block $\CORE_X$ has the following effect:
\begin{center}\small
\begin{tabular}{@{}lll@{}}
\toprule
Input & Register update & Output reader \\
\midrule
$\rd=\M$ & none & $\M$ \\
$\rd=\R$ and $X>0$ & $X\gets X-1$ & $\M$ \\
$\rd=\R$ and $X=0$ & none & $\R$ \\
\bottomrule
\end{tabular}
\end{center}
In every branch all six pile symbols are consumed, the top becomes $\bb{gS}$, and $\R$'s $\bb{yS}$ has been played. All other boundary card types retain their multiplicities.
\end{lemma}
\begin{proof}
Let $c_X$ be $\bb{g2}$ or $\bb{g3}$. The complete traces are
\begin{center}\small
\begin{tabularx}{\linewidth}{@{}p{25mm}X@{}}
\toprule
Branch & Ordered actions \\
\midrule
$\rd=\M$ &
$\M:+c_X$, $\R:+\bb{gS}$, $\R:-\bb{yS}$,
$\R:+\bb{gS}$, $\R:+\bb{g0}$,
$\M:+\bb{gS}$, $\M:+\bb{gS}$.\\
$\rd=\R$, $X>0$ &
$\R:+c_X$, $\M:-U_X$, $\R:-\bb{yS}$,
$\R:+\bb{gS}$, $\R:+\bb{gS}$,
$\R:+\bb{g0}$, $\M:+\bb{gS}$,
$\M:+\bb{gS}$.\\
$\rd=\R$, $X=0$ &
$\R:+c_X$, $\M:+\bb{gS}$, $\M:+\bb{gS}$,
$\M:+\bb{g0}$, $\R:+\bb{gS}$,
$\R:-\bb{yS}$, $\R:+\bb{gS}$.\\
\bottomrule
\end{tabularx}
\end{center}
If the register is nonzero, exactly one $U_X$ is played. Directly after it, $\R$ plays the Skip $\bb{yS}$ and keeps the turn, preventing a second register card from being played, regardless of multiplicity. The other two branches do not touch $U_X$. Every hand play in the displayed traces is of a unique compatible type.
\end{proof}

\subsection{Draining temporary address cards}

\begin{lemma}[Drain gadgets]\label{lem:drain}
Suppose the top is $\bb{g0}$, $H_\R=FJ^q$, and $\M$ has at least $q+1$ copies of $J$, with no other red card or card of symbol $1$ in either hand. If $\rd=\R$, the word $\bb{g1\ rS\ gS}$ removes all $q$ copies of $J$ from $\R$ and exactly $q+1$ from $\M$. If $\rd=\M$ and $q=0$, it removes exactly one $J$ from $\M$. Both cases end with top $\bb{gS}$, $\rd=\R$, and $H_\R=\{F\}$.

Likewise, if the top is $\bb{gS}$, $\rd=\R$, $H_\R=FJ^q$, and $\M$ has at least $q$ copies of $J$ and no competing red or symbol-$1$ card, the word $\bb{rS\ rS\ gS}$ removes precisely $q$ copies of $J$ from each hand, with top $\bb{gS}$, $\rd=\R$, and $H_\R=\{F\}$ at the end.
\end{lemma}
\begin{proof}
In the first case with $\rd=\R$, $\R$ draws and plays $\bb{g1}$ and $\M$ must play one $J$. Then the players alternately play $J$ from their hands until $\R$'s $q$ copies are exhausted. Thus $\M$ has played $q+1$ copies. Next $\R$ draws and plays $\bb{rS}$ and $\bb{gS}$, retaining the turn after both Skips. With $\rd=\M$, $q=0$, $\M$ draws and plays $\bb{g1}$, $\R$ draws and retains the incompatible $\bb{rS}$, $\M$ plays one $J$, and $\R$ plays the retained $\bb{rS}$ and then draws and plays $\bb{gS}$.

For the second gadget $\R$ draws and plays the first $\bb{rS}$. Then each $J$ in $\R$'s hand alternates with a $J$ in $\M$'s hand. When there are no copies left in $\R$'s hand, $\R$ plays the second drawn $\bb{rS}$ and the final $\bb{gS}$. This includes $q=0$, with an empty alternation. Induction on $q$ makes the counting exact. The reserve types exclude premature victory.
\end{proof}

\section{Correctness of the simulation}\label{sec:simulation}

\subsection{Instruction segments}

\begin{proof}[Proof of Lemma~\ref{lem:spec}]
Write $J_M,J_R$ for the numbers of copies of $J$ held by the respective players. Unless explicitly indicated otherwise, the register card types $U_A,U_B$ are unchanged by the intermediate gadgets. The number of reserve cards $E$ remains positive.

\paragraph{Increment segment.}
Let $o=\Off_i(t)$. The active case $j=0$ passes through $\TEST$ without changing $J_M=0$ and without setting a flag. In a passive case $j\ge1$, the test removes two copies of $J$, leaves $J_M=2j-2$, and installs one flag $\bb{r4}$. The block $\PRE(J^oU_X)$ then unconditionally places $o$ copies of $J$ and one copy of $U_X$ in $\M$'s hand. Next $\GIVE$ places a $\bb{yS}$ in $\R$'s hand and $\FLAG$ clears the flag, so that
\[
\begin{array}{c|cc}
&j=0&j\ge1\\\hline
J_M&o&2j-2+o\\
\rd&\M&\R
\end{array}
\qquad (\text{top }\bb{g0}).
\]
The $o$ copies of $\Pi$ add another $o$ copies of $J$ to the current reader's hand: to $\M$ in the active case, to $\R$ in the passive case.

Now consider the next three pile symbols $c_X\,\bb{gS\ gS}$. In the active branch $\M$ draws and plays $c_X$; $\R$ draws and plays $\bb{gS}$, plays its stored $\bb{yS}$, and draws and plays the second $\bb{gS}$. Thus the increment in the number of copies of $U_X$ is preserved. In the passive branch $\R$ draws and plays $c_X$, whereupon $\M$ must play precisely one copy of $U_X$. The preceding $\PRE$ guarantees that such a copy is available. Player $\R$ immediately plays its $\bb{yS}$, followed by both green Skips, canceling the increase caused by loading. In either branch the top is now $\bb{gS}$ and $\rd=\R$.

The closing $\bb{rS\ rS\ gS}$ drains exactly the $o$ garbage $J$ cards from $\R$ and the corresponding $o$ from $\M$ in the passive branch. In the active branch it drains nothing. The resulting values are
\[
\begin{array}{c|cc}
&j=0&j\ge1\\\hline
J_M\text{ at exit}&2o&2j-2\\
X\text{ at exit}&X+1&X.
\end{array}
\]
These are the required transitions.

\paragraph{Conditional decrement segment.}
Let $o,z$ be as in~\eqref{eq:oz}. After the prefix block
\[\TEST\,\PRE(J^{z+1})\,\GIVE\,\FLAG\,\Pi^o\]
we have
\[
\begin{array}{c|cc}
&j=0&j\ge1\\\hline
J_M&z+1+o&2j-2+z+1\\
J_R&0&o\\
\rd&\M&\R
\end{array}
\qquad (\text{top }\bb{g0}).
\]
The next single $\bb{g0}$ swaps the reader. Applying Lemma~\ref{lem:core} to $\CORE_X$ and then applying the following $\bb{g0}$ yields three cases:
\[
\begin{array}{c|ccc}
&j\ge1&j=0,\ X>0&j=0,\ X=0\\\hline
\rd&\R&\R&\M\\
\Delta X&0&-1&0\\
J_M&2j-2+z+1&z+1+o&z+1+o\\
J_R&o&0&0
\end{array}
\qquad (\text{top }\bb{g0}).
\]
Because $z-o\ge0$, the block $\Pi^{z-o}$ adds $z-o$ garbage copies of $J$ to $\R$ in the first two cases, and the same number to $\M$ in the third. The final $\bb{g1\ rS\ gS}$ applies Lemma~\ref{lem:drain}: the first two cases have $\rd=\R$, while the third has $\rd=\M$ and $J_R=0$. The number of copies of $J$ remaining with $\M$ is exactly
\begin{align*}
j\ge1:\quad &(2j-2+z+1)-\bigl(o+(z-o)+1\bigr)=2j-2,\\
j=0,\ X>0:\quad &(z+1+o)-\bigl((z-o)+1\bigr)=2o,\\
j=0,\ X=0:\quad &(z+1+o)+(z-o)-1=2z.
\end{align*}
The register changes are precisely those listed in the preceding table. All cases end in boundary configurations with $H_\R=\{F\}$.

\paragraph{Halting segment.}
In $\HALT$, $\TEST$ leaves top $\bb{gS}$ and $H_\R=\{F\}$ ($\eta=0$), so Lemma~\ref{lem:elementary}(4) applies directly to $\TEST\,\FLAG$, leaving top $\bb{g0}$ with reader $\M$ when $j=0$, and reader $\R$ when $j\ge1$. In the active branch, $\M$ draws and plays the next $\bb{g0}$. Then $\R$ draws and plays $\bb{gS}$ and $\bb{bS}$. Since the top card is now blue, $\R$ immediately plays its sole remaining hand card $F=\bb{b9}$ and wins. No register card is played, and the remaining unused symbols in the halting segment do not need to be consumed.

In the passive branch the complete drawn-card sequence in the tail is
\[
\R:+\bb{g0},\quad
\M:+\bb{gS},\quad
\M:+\bb{bS},\quad
\M:+\bb{gS},\quad
\M:+\bb{g0},\quad
\R:+\bb{gS}.
\]
Player $\R$ never has an opportunity to play $F$. The hands return to the canonical types with $J_M=2j-2$, $H_\R=\{F\}$ and unchanged registers, and the final top is $\bb{gS}$ with reader $\R$.

\paragraph{Uniqueness of the playable type.}
The complete local traces establish this claim. The only red card normally retained by $\M$ is $J$; its flag exists only between $\TEST$ and $\FLAG$, where no red top is presented to that player. Both yellow register types may coexist, but whenever $\M$ plays a yellow card in a decrementing step, the next turn belongs to $\R$, whose single $\bb{yS}$ is immediately played and retains the turn. Lemmas~\ref{lem:elementary}--\ref{lem:drain} and the halting trace cover the remaining cases. Thus no tie-breaking is used along the greedy execution.
\end{proof}

\subsection{Initialization and machine simulation}

\begin{proof}[Proof of Theorem~\ref{thm:main}]
Fix the length-$17$ prefix
\begin{equation}\label{eq:prefix}
u=\bb{y8\ b9}\,(\bb{g5\ g5})^6\,\bb{g5\ g0\ gS}.
\end{equation}
The first fourteen cards are alternately dealt, giving $\M$ the reserve card $E$ and six $\bb{g5}$, and $\R$ the card $F$ and six $\bb{g5}$. The fifteenth card is the ordinary top discard $\bb{g5}$. The two players alternately discard the twelve $\bb{g5}$ in their hands; every such play is of a unique compatible type. Then $\M$ draws and plays $\bb{g0}$ and $\R$ draws and plays $\bb{gS}$. This establishes $\mathcal C(0,0,0)$ immediately before the first segment $S_0$.

As illustrated in Figure~\ref{fig:jump}, the program counter is distributed over the segment index and the jump counter. At the entrance of segment $S_i$, let the boundary state be $\mathcal C(j,a,b)$ and define
\[
p\equiv i+j\pmod N.
\]
We prove inductively over completed segments that $p$ is the instruction label currently simulated by the machine, and that $a,b$ are its register values. The initial state has $i=j=0$. When $j>0$, Lemma~\ref{lem:spec} turns the segment into a passive one, leaving the register values unchanged and replacing $j$ by $j-1$. Advancing the segment index to $i+1$ modulo $N$ leaves $(i+j)\bmod N$ unchanged. When $j=0$, the segment implements the instruction $I_i$ as specified. Any nonhalting transition to label $t$ leaves a new counter $j'=\Off_i(t)$, or $j'=z\equiv\Off_i(t)\pmod N$ for the zero branch. Therefore
\[
(i+1)+j'\equiv t\pmod N,
\]
and the stored register values evolve exactly as in the Minsky machine. Since $0\le j'\le2N-1$, a finite number of passive segments separate successive simulated machine instructions.

At an active $\HALT$, $\R$ wins by Lemma~\ref{lem:spec}. Before then $\R$ retains $F$ and $\M$ retains at least one $E$, so neither player can win prematurely (see also Appendix~\ref{app:semantics}). If the machine never halts, infinitely many segments are entered and the play is infinite. If it halts, the last $U_A,U_B$ counts agree with the machine's registers by the induction. The uniqueness-of-type claim also follows from Lemma~\ref{lem:spec}.

The gadget lengths are linear in $o,z=O(N)$, and each program has $N$ segments. Hence $|w|=O(N^2)$; Appendix~\ref{app:size} gives an explicit upper bound.
\end{proof}

\section{Maximal finite play duration}\label{sec:duration}

Let $T(uw^\omega)$ be the number of complete player turns until termination when the play terminates, counting turns on which the player only draws and keeps a card. Different representations of the same infinite word are allowed. For an input description $(u,w)$ define its length to be $|u|+|w|$.

\begin{definition}
For $n\in\mathbb N$, let
\[
\UU(n)=\max\Bigl(\{0\}\cup
\bigl\{T(uw^\omega): |u|+|w|\le n,\ w\ne\varepsilon,
\ \text{the greedy play terminates}\bigr\}\Bigr).
\]
\end{definition}

There are only finitely many pairs of finite words over $\Gamma$ of total length at most $n$, so the maximum is well defined and finite. The next proposition shows that it has no total computable upper bound.

\begin{proposition}\label{prop:bb}
No total computable function $f:\mathbb N\to\mathbb N$ is an upper bound for $\UU$ at every argument.
\end{proposition}
\begin{proof}
Suppose $f(n)\ge\UU(n)$ for every $n$. Given a two-counter machine $\mathcal M$, construct $u w_{\mathcal M}^\omega$ as in Theorem~\ref{thm:main} and compute $n=|u|+|w_{\mathcal M}|$. Simulate the greedy play for $f(n)+1$ turns unless it terminates earlier. If it terminates, the machine halts. Otherwise it cannot ever terminate, since the duration of any terminating play with description length at most $n$ would not exceed $f(n)$. This decides the halting problem, a contradiction.
\end{proof}

\section{Unrestricted play and existential termination}\label{sec:unrestricted}

The uniqueness and simulation claims of Theorem~\ref{thm:main} require mandatory compatible play. We now allow all legal actions from Section~\ref{sec:model}, including drawing instead of playing an available compatible card.

\subsection{Infinite legal plays}

\begin{proposition}\label{prop:stall}
Under unrestricted legal play, an infinite play exists from every initial state with an infinite draw pile and two nonempty hands.
\end{proposition}
\begin{proof}
Every player may choose to draw instead of playing an existing compatible card, and may retain the drawn card even when it is compatible. If both players always do this, no card is ever played, no hand becomes empty, and the game continues forever.
\end{proof}

Thus the property that every legal play terminates is false for all such inputs. Even existence of a finite play cannot be inferred from the greedy simulation, as the next example shows.

\subsection{A passive halting segment under unrestricted play}

\begin{proposition}[Failure of the passive halting gadget]\label{prop:passive-halt}
Under unrestricted legal play, a passive $\HALT$ segment of~\eqref{eq:halt} can end in a victory by $\R$. Hence the conclusion of Lemma~\ref{lem:spec} does not extend to arbitrary legal play.
\end{proposition}
\begin{proof}
Start in $\mathcal C(1,0,0)$. After $\TEST\,\FLAG$, the top is $\bb{g0}$, the next reader is $\R$, the hands are $H_\M=E^+$ and $H_\R=\{F\}$, and the remaining segment is
\[
\bb{g0\ gS\ bS\ gS\ g0\ gS}.
\]
Player $\R$ draws and plays $\bb{g0}$. Player $\M$ draws the compatible $\bb{gS}$ but retains it. Player $\R$ draws and retains $\bb{bS}$, which is incompatible with $\bb{g0}$. Now $\M$ plays the stored $\bb{gS}$, draws and plays the next $\bb{gS}$ and $\bb{g0}$, and passes the turn. Player $\R$ draws and plays the last $\bb{gS}$, then plays the stored $\bb{bS}$ and finally the remaining $\bb{b9}$. Both Skips retain $\R$'s turn, and the last play empties $\R$'s hand. The passive segment therefore terminates under these legal choices.
\end{proof}

\subsection{Existential termination via VASS reachability}

The existential problem nevertheless has a decision procedure independent of the greedy construction.

\begin{definition}[Existential termination]
Given a finite description $(u,w)$ of an ultimately periodic draw pile,
\textsc{Uno-Exists-Finite} asks whether some finite legal play exists
under the rules of Section~\ref{sec:model}, without imposing $\sigma$.
For $P\in\{\M,\R\}$, \textsc{Uno-Exists-$P$-Win} asks whether
such a play can end with $P$ winning. The initial deal and draw-pile
order are fixed by the input; only the players' legal choices vary.
\end{definition}

\begin{theorem}[Decidability and complexity of existential termination]\label{thm:vass}
\textnormal{\textsc{Uno-Exists-Finite}} and \textnormal{\textsc{Uno-Exists-$P$-Win}} are decidable---and, more precisely, primitive recursive---for ultimately periodic infinite draw piles under the rules of Section~\ref{sec:model}.
\end{theorem}
\begin{proof}
\textbf{Counters and finite control.}
We construct an effectively presented vector addition system with states (VASS),
whose reachability relation captures the legal nonterminal histories. Its counter vector is
$h=(h_{P,x})_{P\in\{\M,\R\},x\in\Gamma}\in\mathbb N^{2|\Gamma|}$;
$h_{P,x}$ is the multiplicity of type $x$ in $P$'s hand. The finite control records
the current player, top discard, the finite-prefix/period position of the next deck
symbol, and a finite phase specifying ordinary turns, a just-drawn card, or a pending
forced two-card draw. The pile position has only $|u|+|w|$ possibilities:
a pointer in $u$ moves forward and a pointer in $w$ moves cyclically.

\smallskip\noindent\textbf{Legal moves.}
The rules are expressed by a finite collection of transitions, each enabled solely by
nonnegativity of the updated vector. Playing a compatible type $x$ from the hand
subtracts $e_{P,x}$ and changes finite control according to its symbol; this transition
is enabled exactly when $h_{P,x}\ge1$. Drawing the next fixed deck symbol $x$ adds
$e_{P,x}$ and advances the pointer, regardless of the current hand: \emph{voluntary
drawing is always legal}. When the newly drawn $x$ is compatible with the previous
top, the player may keep it or play \emph{that same} $x$ immediately. In the latter
case the addition and subtraction of $e_{P,x}$ cancel, but the deck pointer advances
and the played-card effects apply. Incompatible drawn cards must be kept. A Draw Two
card is simulated by two forced counter increments and a skipped turn; a Skip card
only changes finite control. Compatibility is tested between card types stored in
finite control, not by a predicate on the counters. Thus \emph{no test for the absence
of a playable hand card} is needed: the player may always draw.

\smallskip\noindent\textbf{Nonterminal histories.}
The game ends immediately when a hand becomes empty.
On every ordinary (nonterminal) discard transition, after subtracting the discarded
card we require that the acting player's hand remains nonempty. This is a finite
disjunction of \emph{positive} counter tests. A test
$h_{P,y}\ge1$ can be realized within a VASS by temporarily subtracting and restoring
$e_{P,y}$ in two transitions through a fresh intermediate control state. Hence it does
not introduce a zero test. All ordinary draw transitions preserve nonempty hands (including immediate play of a drawn card, which leaves hand counts unchanged), and
all initial hands are nonempty. Any counter path through these nonterminal controls
therefore represents a legal unfinished game.

\smallskip\noindent\textbf{Terminal histories.}
For each player $P$, add a separate \emph{terminal} transition for playing a compatible
card from that player's hand. This transition performs the subtraction and enters a
terminal control state with no ordinary game transitions; it does not assert that the
hand is empty. Such a transition is accepted \emph{only if} every counter of the
winning player is exactly zero in the terminal state. To reduce this partial-zero
reachability condition to ordinary VASS reachability, add, in the terminal state,
self-loops subtracting one copy of each card type held by the \emph{other} player,
and ask to reach the all-zero marking in a designated final control state.
No counters of $P$ can change after the terminal transition, so the all-zero marking
is reachable exactly if that discard truly emptied $P$'s hand. A path witnessing
reachability projects to a legal game ending with $P$'s victory; conversely, any
such game produces a witnessing VASS path. Taking the disjunction over both players
gives finite-play existence.

Ordinary reachability for VASS (equivalently, Petri nets with finite control) is
decidable~\cite{mayr1984,kosaraju1982}. Moreover, the constructed VASS has fixed
dimension $d=2|\Gamma|=96$ independent of $(u,w)$, with $|u|+|w|$ entering only the
size of the finite control. By the fixed-dimension upper bound of Leroux and
Schmitz~\cite{lerouxschmitz2019}, reachability in $d$-dimensional VASS is primitive
recursive (lying in the fast-growing complexity class $\mathcal F_{d+4}=\mathcal F_{100}$);
therefore both decision problems are primitive recursive.
\end{proof}

\begin{remark}
The theorem is about \emph{existence of some terminating play} under voluntary-draw
rules, not the existence of a winning strategy against an opponent who may also choose
whether to draw voluntarily (contrast Corollary~\ref{cor:forced} for forced-play rules).
The voluntary-draw two-player strategy problem has alternating quantifiers and does not
follow from VASS reachability; its decidability remains open, as does the exact
elementary complexity of \textsc{Uno-Exists-Finite}. The result also relies on our
explicitly specified card alphabet without Wild Draw Four's hand-color restriction,
which would require a separate analysis.
\end{remark}

\section{Conclusion}\label{sec:conclusion}
Color-or-symbol matching, Skip, and Draw Two suffice to simulate
two-counter machines under mandatory compatible play. The ultimately
periodic draw pile has a finite description; unbounded card
multiplicities in the hands provide the memory.

Greedy termination, designated-player victory, and winning-strategy
existence under forced-play rules are $\Sigma^0_1$-complete, whereas
existence of a finite legal play under official voluntary-draw rules is
primitive recursive via fixed-dimension VASS reachability. Mandatory play
supplies the zero test used by the simulation; voluntary drawing permits
the existential problem to be represented without zero tests. Determining
the exact complexity of \textsc{Uno-Exists-Finite} and the decidability of
winning strategies under voluntary drawing remain natural open questions.

\appendix
\section{Operational semantics and nontermination safety}\label{app:semantics}

\begin{definition}[Microstep and read boundary]
A full state consists of both finite hand multisets, the top discard, the player whose turn it is, and the next unread draw-pile index. A \emph{microstep} is a play from the hand, an ordinary one-card draw with its optional immediate play, or a compulsory two-card draw caused by $D$. The last operation consumes two consecutive draw-pile cards and grants no turn to their recipient. A \emph{read boundary}, as introduced in Section~\ref{sec:encoding}, occurs after all hand plays prescribed by $\sigma$ have been completed and immediately before the next draw-pile consumption. The reader $\rd$ is the player who will consume the next card. Only finitely many hand plays can precede that draw, because each reduces the total number of cards held.
\end{definition}

\begin{lemma}[No early victory]
Before an active $\HALT$, none of the segments in the construction can terminate the game. At every segment boundary $\R$ has its persistent card $F=\bb{b9}$, while $\M$ retains at least one $E=\bb{y8}$. The same cards remain unplayed during every nonterminal gadget, although either hand may temporarily contain other cards.
\end{lemma}
\begin{proof}
The exact playable types are exhausted by Lemmas~\ref{lem:elementary}--\ref{lem:drain} and the halting trace. Neither $F$ nor $E$ is played there; the $\PRE$ gadget may create additional $E$ but never consumes it. Only the active halting tail reaches a blue top at a turn at which $\R$ can play $F$. Thus the persistent reserves prevent either hand from emptying prematurely.
\end{proof}

\section{Transition certificates at block boundaries}\label{app:certificates}

The table summarizes \emph{boundary transformations}. Each row inherits the full input hypotheses of its corresponding lemma in Section~\ref{sec:gadgets}; the middle column lists the conditions most relevant to the transformation. Here $q$ is the number of temporary copies of $J$ held by $\R$. Complete action traces appear in the proofs and in Appendix~\ref{app:traces}.

\begin{center}\small
\begin{tabularx}{\linewidth}{@{}p{25mm}p{48mm}X@{}}
\toprule
Block & Relevant input & Output transformation \\
\midrule
$\TEST$ & $\mathcal C(0,a,b)$ or $\mathcal C(j,a,b)$, $j>0$ &
$J_M:0\mapsto0$ or $2j\mapsto2j-2$ with flag $\bb{r4}$; top $\bb{gS}$, $\rd=\R$\\
$\PRE(x_1\cdots x_m)$ & top $\bb{gS}$, $\rd=\R$, $H_\R=\{F\}$ &
Add $x_1,\ldots,x_m$ and possibly $E$ to $H_\M$; same top and reader\\
$\GIVE$ & top $\bb{gS}$, $\rd=\R$ &
Add $\bb{yS}$ to $H_\R$; top $\bb{g0}$, $\rd=\R$\\
$\FLAG$ & top $\bb{g0}$ (or $\bb{gS}$, $\eta=0$); Lemma~\ref{lem:elementary}\textup{(4)} &
Remove flag; top $\bb{g0}$; $\rd=\R$ if flag present, $\M$ otherwise\\
$\Pi$ & top $\bb{g0}$, $\rd=P$ &
$J_P\gets J_P+1$; reader and top unchanged\\
$\CORE_X$ & top $\bb{g0}$, $\bb{yS}\in H_\R$ &
Reader $\M\to\M$ (no update); $\R\to\M$ when $X>0$ (decrement); $\R\to\R$ when $X=0$\\
$\bb{g1\ rS\ gS}$ & top $\bb{g0}$, $\rd=\R$, $J_R=q$, $J_M\ge q+1$ &
$J_R\gets0$, $J_M\gets J_M-q-1$; top $\bb{gS}$, $\rd=\R$\\
$\bb{rS\ rS\ gS}$ & top $\bb{gS}$, $\rd=\R$, $J_R=q$, $J_M\ge q$ &
$J_R\gets0$, $J_M\gets J_M-q$; top $\bb{gS}$, $\rd=\R$\\
\bottomrule
\end{tabularx}
\end{center}

\section{Additional local transition traces}\label{app:traces}
This appendix collects parameterized traces supplementing
Lemmas~\ref{lem:elementary}--\ref{lem:drain}. The notations
$P:+x$, $P:\langle x\rangle$, $P:-x$ mean respectively draw-and-play, draw-and-keep,
and play-from-hand. Each line presupposes a boundary state with no green or
zero-symbol card in either hand and no additional compatible red hand type.

\paragraph{The one-card switch and the window.}
With top $\bb{g0}$, $\rd=P$, and both hands restricted as above,
the word $\bb{g0}$ is consumed as $P:+\bb{g0}$, after which the
opponent is the next reader. For $\Pi=\bb{r1\ g0}$, the trace is
$P:\langle J\rangle$ followed by
$\overline P:+\bb{g0}$. Thus the next reader is again $P$,
and $P$ holds one additional $J$. The trace applies to either reader
and proves both cases of Lemma~\ref{lem:elementary}(5).
At a read boundary with top $\bb{gS}$, a single $\bb{g0}$ is also
compatible and performs the same reader swap.

\paragraph{Drain with an arbitrary garbage multiplicity.}
In the $\bb{g1\ rS\ gS}$ case with reader $\R$, first
$\R:+\bb{g1}$ and $\M:-J$. The subsequent $2q$ hand plays are
\[
(\R:-J,\ \M:-J)^q;
\]
then $\R:+\bb{rS}$ and $\R:+\bb{gS}$. Player $\R$
has played $q$ copies of $J$, and $\M$ has played $q+1$.
If $q=0$ the middle block is empty and the initial $\M:-J$
is still required; this is the reason for the $q+1$ precondition.
When the reader is $\M$ and $q=0$, the trace is
$\M:+\bb{g1}$, $\R:\langle\bb{rS}\rangle$, $\M:-J$,
$\R:-\bb{rS}$, $\R:+\bb{gS}$. For the second drain gadget,
$\bb{rS\ rS\ gS}$ with reader $\R$, the trace is
$\R:+\bb{rS}$, followed by $(\R:-J,\ \M:-J)^q$,
then $\R:+\bb{rS}$ and $\R:+\bb{gS}$. In particular,
$q=0$ involves only the three draw-and-play actions. These traces
establish both claims for every $q\in\mathbb N$.

\paragraph{The increment tail.}
Immediately before the tail
$c_X\,\bb{gS\ gS\ rS\ rS\ gS}$, the reader is $\M$
in the active branch and $\R$ in the passive branch, and
$H_\R=F\bb{yS}$ in the active branch and
$H_\R=FJ^o\bb{yS}$ in the passive branch.
In the active branch the trace through
$c_X\,\bb{gS\ gS}$ is
$\M:+c_X$, $\R:+\bb{gS}$, $\R:-\bb{yS}$, $\R:+\bb{gS}$;
there is no register decrement, and the second drain gadget with
$q=0$ completes the segment. In the passive branch, the trace is
$\R:+c_X$, $\M:-U_X$, $\R:-\bb{yS}$,
$\R:+\bb{gS}$, $\R:+\bb{gS}$;
then the second drain gadget removes precisely $o$ copies of $J$
from each player. The preceding $\PRE$ guarantees that a copy of
$U_X$ is available; playing one copy cancels the increase caused by
loading, independently of the original register value.

\paragraph{Exhaustion of the cases.}
At each block boundary the hands have only the types explicitly
listed in the corresponding input hypotheses. Inside $\TEST$ the only
additional hand types are $\bb{g0}$, $\bb{rS}$, and the eventual flag
$\bb{r4}$, already accounted for in the full test trace; $\PRE$
does not grant a turn to $\M$; $\GIVE$, $\FLAG$, and the $\Pi$
blocks are covered above and in Lemma~\ref{lem:elementary}; the three
cases of $\CORE_X$ are exhausted by Lemma~\ref{lem:core}; and the two
drain types and their degeneracies are enumerated here. Thus there
are no omitted value-dependent branches in the segment proof. This
case-exhaustion argument is independent of the finite regression tests.

\section{Explicit description-length bound and example}\label{app:size}

The block $\PRE$ has length $3\lceil m/2\rceil+1$ for a list of $m$ symbols. Directly from~\eqref{eq:inc}--\eqref{eq:halt}, the length of an instruction segment is
\begin{align*}
|S_i|&=22+2o+3\left\lceil\frac{o+1}{2}\right\rceil &&\text{for $\INC$},\\
|S_i|&=27+2z+3\left\lceil\frac{z+1}{2}\right\rceil &&\text{for $\DEC$},\\
|S_i|&=17 &&\text{for $\HALT$}.
\end{align*}
As $o\le N-1$ and $z\le2N-1$, all cases satisfy $|S_i|\le25+7N$, and therefore
\[
|w|\le N(25+7N),\qquad |u|=17.
\]

For instance, take the two-instruction machine $I_0=\INC(A,1)$, $I_1=\HALT$. Here $N=2$, $\Off_0(1)=0$, the increment segment has length $25$, and the halting segment has length $17$, so that $|w|=42$. After execution of the first segment $\M$ has exactly one $\bb{y2}$; the next segment is active and $\R$ wins with the register value unchanged.

\section{Computational checks and their scope}\label{app:checks}

The supplementary scripts \texttt{verify\_gadgets.py} and \texttt{uno\_program\_tests.py} implement a stepwise simulator, a gadget compiler, and regression tests. The reported Python~3 runs comprise $173\,250$ parameterized $\INC$/$\DEC$ segment checks ($10\,500$ for $\INC$ and $162\,750$ for $\DEC$), $525$ $\HALT$-segment checks, $19\,984$ elementary-block checks, and independent program suites, including $400$ pseudorandom two-counter programs generated with seed $2331$ (alongside $7$ deterministic smoke cases). Of those $400$ programs, $90$ reached $\HALT$; the other $310$ were followed up to a limit of $100$ segments. Reaching this limit does not establish nontermination.

The scripts are distributed as ancillary files accompanying this preprint. These finite checks supplement the action-trace proofs; they do not establish the universal lemmas on their own.

\section*{Use of artificial intelligence}
In preparing this preprint, the authors used OpenAI ChatGPT (GPT-6 Sol) for editing and exposition, auditing parts of the proofs, translation, discussing the VASS formulation, and writing auxiliary verification scripts, and Anthropic Claude (Opus 5.5) for comparative review and editorial suggestions. These tools are not authors. The authors verified all mathematical claims, proofs, references, and code, and take full responsibility for the originality and accuracy of the content.


\begin{thebibliography}{12}
\bibitem{minsky1967}
Minsky, M.~L. \emph{Computation: Finite and Infinite Machines}. Prentice--Hall, 1967.

\bibitem{demaine2014}
Demaine, E.~D., Demaine, M.~L., Harvey, N.~J.~A., Uehara, R., Uno, T., and Uno, Y.
UNO is hard, even for a single player. \emph{Theoretical Computer Science} \textbf{521} (2014), 51--61.
\href{https://doi.org/10.1016/j.tcs.2013.11.023}{doi:10.1016/j.tcs.2013.11.023}.

\bibitem{churchill2020}
Churchill, A., Biderman, S., and Herrick, A.
Magic: The Gathering Is Turing Complete. In \emph{Proc.\ 10th Int.\ Conf.\ on Fun with Algorithms (FUN 2021)}, LIPIcs \textbf{157}, art.~9 (2020), 9:1--9:19.
\href{https://doi.org/10.4230/LIPIcs.FUN.2021.9}{doi:10.4230/LIPIcs.FUN.2021.9}.

\bibitem{biderman2020}
Biderman, S. Magic: the Gathering is as Hard as Arithmetic.
\emph{arXiv:2003.05119} (2020). \url{https://arxiv.org/abs/2003.05119}.

\bibitem{yugioh2026}
Nicolosi, O., Pisciotta, F., and Bresolin, L.
Deciding winning strategies in Yu-Gi-Oh!\ TCG is hard. \emph{arXiv:2603.02863} (2026).
\url{https://arxiv.org/abs/2603.02863}.

\bibitem{mayr1984}
Mayr, E.~W. An algorithm for the general Petri net reachability problem.
\emph{SIAM Journal on Computing} \textbf{13}(3) (1984), 441--460.
\href{https://doi.org/10.1137/0213029}{doi:10.1137/0213029}.

\bibitem{kosaraju1982}
Kosaraju, S.~R. Decidability of reachability in vector addition systems (preliminary version).
In \emph{Proc.\ 14th Annu.\ ACM Symp.\ on Theory of Computing (STOC '82)} (1982), 267--281.
\href{https://doi.org/10.1145/800070.802201}{doi:10.1145/800070.802201}.

\bibitem{lerouxschmitz2019}
Leroux, J., and Schmitz, S.
Reachability in vector addition systems is primitive-recursive in fixed dimension.
In \emph{Proc.\ 34th Annu.\ ACM/IEEE Symp.\ on Logic in Computer Science (LICS 2019)} (2019), 1--13; \emph{arXiv:1903.08575}.
\url{https://arxiv.org/abs/1903.08575}.

\bibitem{mattelrules}
Mattel. \emph{UNO Deluxe: Instruction Sheet}, M2062-0920 (2007).
\url{https://service.mattel.com/instruction_sheets/M2062-0920.pdf}.
\end{thebibliography}
\end{document}